\documentclass[12pt]{article}

\begin{document}

\sloppy
\begin{flushright}{SIT-HEP/TM-12}
\end{flushright}
\vskip 1.5 truecm
\centerline{\large{\bf F-term, D-term and hybrid brane inflation}}
\vskip .75 truecm
\centerline{\bf Tomohiro Matsuda
\footnote{matsuda@sit.ac.jp}}
\vskip .4 truecm
\centerline {\it Laboratory of Physics, Saitama Institute of
 Technology,}
\centerline {\it Fusaiji, Okabe-machi, Saitama 369-0293, 
Japan}
\vskip 1. truecm
\makeatletter
\@addtoreset{equation}{section}
\def\theequation{\thesection.\arabic{equation}}
\makeatother
\vskip 1. truecm

\begin{abstract}
\hspace*{\parindent}
We study inflation and reheating in models for the brane universe,
considering hybrid brane inflation without tachyon condensation.
We expect that some fields that are localized on different branes 
interact with $O(1)$ coupling when branes are on top of each other, 
while the interaction vanishes when branes are separated at a distance.
If the interaction is needed to avoid spontaneous breaking of 
supersymmetry on the brane, our idea for hybrid brane inflation works.
In our model, when branes are far apart, supersymmetry is 
spontaneously broken by the
fields on a brane, which induces inflation.
The inflaton field is the moduli for the brane distance.
At the end of inflation, when branes come close, supersymmetry
is restored by the interaction between fields on the
branes, then the oscillation starts to reheat the Universe.
In this paper we construct explicit models for F-term and D-term
 inflation.
There are at least two major advantages.
The most attractive point is that reheating is natural in our model,
because the trigger field is not the tachyon but a conventional 
field on the brane.
The serious constraint from the loop correction, which always appears in
 conventional models for hybrid inflation, is avoided.
\end{abstract}

\newpage
\section{Introduction}
\hspace*{\parindent}
Although the quantum field theory achieved great successes,
consistent scenario that includes quantum gravity is still lacking.
String theory would be the most promising scenario in this direction.
The requirement of additional dimensions is a characteristic feature of
the string theory, which ensures the required consistency.
At first, sizes of extra dimensions had been assumed to be
as small as $M_p^{-1}$.
Then it is shown that there is no reason to believe such tiny
compactification radius\cite{Extra_1}. 
In models with large extra dimensions, 
the observed Planck mass is obtained by the relation $M_p^2=M^{n+2}_{*}V_n$,
where $M_{*}$ and $V_n$ denote the fundamental scale of gravity
and the volume of the $n$-dimensional compact space.
If one assumes more than two extra dimensions, $M_{*}$ may be
close to the TeV scale without conflicting any observable bound.
The most natural embedding of this picture in
the string theory context will be realized by the branes.
Of course, the world on the branes will be a viable candidate for the
Universe even if the fundamental scale is not so low as the TeV scale.
In the brane world scenario, there is no obvious reason to 
believe that the fundamental scale is as high as the Planck scale.

Although the idea of the brane world inspired us to construct new types of
phenomenology, a drastic modification is needed
for the conventional cosmological scenarios.
Models of inflation and baryogenesis\cite{low-baryon} are especially
sensitive to such a low fundamental scale, i.e., $M_* << M_{GUT}$ where
$M_{GUT}$ denotes the standard (old) GUT scale. 
To avoid extreme fine-tunings, one should reconstruct
conventional scenarios of the standard cosmology.
This requires the inclusion of novel ideas that are quite different from the
conventional ones.
For example, if one puts the inflaton field on the brane, their masses
are required to be unnaturally small\cite{fine-tune}.
On the other hand, in generic cases, the mass of the inflaton is bounded
from below to achieve successful reheating. 
Thus it seems quite difficult to construct a model for
inflation driven by a field on the brane.
\footnote{
A way to avoid this difficulty is put forward by Arkani-Hamed et
al.\cite{Arkani-inflation}, where inflation is assumed to occur 
before the stabilization of the internal dimensions.
In this case, however, the late oscillation of the radion field is a serious 
problem, which may or may not be solved by the second weak
inflation\cite{halyo}.
Alternatively, one may assume that the extra dimensions are stabilized
by some dynamical mechanisms before the Universe exited from
inflation\cite{Mohapatra-1, 
Mazumdar-Bulk-Inflation,matsuda_bulk_inflation}.} 

In the scenario of the brane world, one may find another possibility.
In the scenario of brane inflation\cite{brane-inflation}, 
the branes are displaced from the stable point at the beginning of
inflation, and the interbrane distance is used for the inflaton field.  
In this case, because of the tachyon instability, unconventional 
hybrid inflation is naturally obtained. 
The system develops tachyon modes when the brane distance becomes small,
then leads to a natural end of inflation via the extra trigger field.
This type of scenario has been discussed 
within various settings\cite{tachyonic_brane_inflation, angle_inflation}.
On the other hand, however, there is a serious problem
related to the peculiar properties of the tachyon.
Because the tachyon cannot oscillate after inflation, reheating is not so
easy as the conventional hybrid inflation\cite{tachyon_reheating}.
Thus, it seems very interesting to
construct models for hybrid brane inflation where 
the physics related to the tachyon condensation does not appear.

In this paper we study inflation and the reheating in models for the
brane world, 
considering hybrid brane inflation without 
tachyon\cite{matsuda_brane}.
In any models for the brane universe, it is natural to think that some
fields are localized on branes.
It is also natural to expect that fields on the different branes may have
$O(1)$ couplings when branes are on top of each other, while such
 interactions may vanish when branes are located at a distance.
If the interaction that is needed to avoid spontaneous breaking of
supersymmetry
on the brane vanishes at the beginning of inflation,
our idea for hybrid brane inflation works.
In our model, when branes are far apart, supersymmetry is spontaneously 
broken by the
 field on a brane and the vacuum energy induces inflation.
At the end of inflation, instability is induced by another
field on another brane.
During inflation, their cross terms are suppressed by the
brane distance.
Then at the end of inflation, the cross terms become as large as O(1),
and the field rolls down to the supersymmetric vacuum.
The inflaton is the moduli for the brane distance.
At the end of inflation, when branes come close, supersymmetry
is restored by the interaction terms between fields on the
branes, then the field oscillates to reheat the Universe.
We construct explicit models for F-term and D-term inflation.
There are two advantages compared to the previous models for 
tachyonic brane inflation or standard hybrid inflation.
The most attractive point is that reheating is natural in our model.
The serious constraint from the loop correction is also avoided. 
In our model, unlike other models for hybrid inflation in the brane
Universe, tachyon is not required.
Inflation ends because the localized fields on each brane
begin to interact.
Then the interaction destabilizes the potential on the brane.

In section 2, we construct brane inflation with the F-term.
The model can be used for the secondary weak inflation, however it
cannot be used for the first inflation in generic situations.
In section 3, we show that inflation with the D-term is possible within our
settings.
Unlike conventional D-term inflation in supergravity, the loop
correction is not a serious problem.
No fine-tuning is required for the coupling constants.

\section{F-term inflation with moving brane}
\hspace*{\parindent}
In this section we consider a ``toy'' model where the F-term on
a brane induces inflation.
As in the conventional models for brane inflation, two branes are
needed.
At the beginning of inflation, these two branes are assumed to be 
located at a distance in the extra dimensions.
To make inflation,
at least two fields are required to be localized on each
branes.\footnote{Here we do not specify the mechanism for the localization.}
On one brane, a localized field $S$ is expected to form a superpotential
of the form 
\begin{equation}
W_1=S\Lambda_1^2 
\end{equation}
if the charge of the superfield $S$ under $U(1)_R$ symmetry is the same as
the assigned charge of the superpotential. 
We think one can easily understand that $W_1$ breaks supersymmetry by
the F-term on the brane at $r=r_1$. 
Here $\vec{r_1}$ denotes the location of the corresponding brane in the extra
dimensions.
On the other brane, a superfield $\Phi$ is localized at
$\vec{r}=\vec{r_2}$, with superpotential $W_2=0$.
Here we have assumed that the $U(1)_R$ charge
of the superfield $\Phi$ is 0.\footnote{For simplicity, here we do not
consider other fields that may 
have $U(1)_R$ charges.}
However, when two branes come close at the end of inflation,
the localized fields $S$ and $\Phi$ may interact.
Then the following superpotential will appear on the brane, 
\begin{equation}
W_{1+2}=\lambda S(\Lambda_1^2-\Phi^2).
\end{equation}
More precisely, the interaction depends on the overlap of the
wavefunctions of the fields, which will have the following form
\begin{equation}
\label{1+2}
W_{1+2}=\lambda S(\Lambda_1^2-\Phi^2 \lambda' 
e^{-(M_0|\vec{r_1}-\vec{r_2}|)^2}),
\end{equation}
where $\lambda, \lambda'$ are dimensionless constants, and $M_0^{-1}$
is the width of the wavefunctions.
One may explain each form of the superpotential by imposing continuous
or discrete R-symmetry.
Inflation starts when two branes are located at a distance.
In this case, the inflaton is the moduli that parametrizes the distance
between branes.
The Moduli is flat when supersymmetry is maintained, but is lifted by the
supersymmetry breaking on the brane.
In the low energy effective description,
the situation is similar to the conventional F-term inflaton in
conventional models of supergravity. 
Without non-trivial requirement from the symmetry of the model,
the effective mass of the inflaton would be as large as the Hubble
parameter $H$, which makes it difficult to achieve the
requirement from the 
conventional scenarios of inflation.
Although a successful inflation may be achieved by the additional
symmetries of the brane world, at this time we have no concrete example
where the effective mass of the inflaton is well suppressed for 
F-term inflation. 
Although the model for inflation with the F-term does not seem to be
suitable for the 
first inflation, it is sometimes useful for weak 
inflation\cite{matsuda_brane}.

\section{D-term inflation with moving brane}
\hspace*{\parindent}
As we have discussed in the previous section, F-term inflation suffers 
from the old serious difficulty even if it is extended to the models of 
brane inflation. 
In the past, the idea of D-term inflation was invoked to solve the
problem of F-term inflation in conventional supergravity.
In this section we examine whether
 one can use the same idea to solve the problem 
of brane inflation in the brane world.

Our model for inflation may seem to be a simple modification of the
conventional hybrid inflation.
However, we will discuss the crucial differences in the followings.
As is discussed in ref.\cite{lyth}, loop corrections are the serious
problem for models of the conventional hybrid inflation.
The situation is not changed in the {\it conventional} D-term inflation.
The key difference in our model is that the trigger field on
the brane is placed at the false vacuum {\it without} the problematic
large coupling.

Here we consider a localized Fayet-Iliopoulos term on a brane at
$\vec{r}=0$
 of the form
\begin{equation}
\label{FI}
\xi D \delta(\vec{r})
\end{equation}
where $D$ is an auxiliary field of the vector superfield.
We consider an additional abelian gauge group $U(1)_X$ in the bulk,
while the Fayet-Iliopoulos term for $U(1)_X$ is localized on a brane. 
We also include the fields $\phi_X$ that has $U(1)_X$ charge and
localized on the other brane at $\vec{r}=\vec{r_1}$.
When two branes are located at a distance, $|\vec{r_1}| >>M_*^{-1}$,
the Fayet-Iliopoulos term (\ref{FI}) breaks supersymmetry on the brane
and inflation starts.
\footnote{Here we temporally ignore the derivative terms. }
In this case, as in the conventional models for brane inflation, 
the inflaton field is the moduli that parametrizes the
brane distance.
The moduli is denoted by $\sigma=M_*^2 r_1$, 
where $M_*$ is the fundamental scale of the model. 
As we are considering D-term inflation, the mass of the inflaton
($m_{\sigma}$) may be much smaller than the Hubble parameter.
Then a modest limit is $m_{\sigma} \ge m_{3/2}$,
where  $m_{3/2}$ is the gravitino mass in the true vacuum.

\underline{Effect of the derivative terms}

Here we consider the simplest example, five-dimensional theory that is
made chiral by choosing the right boundary conditions\cite{orbifold5d}.
An abelian gauge multiplet of the five-dimensional gauge sector consists
of a vector superfield $V$ whose components are the four-dimensional
part of the vector gauge field $A^{\mu}$, the left-handed gaugino,
auxiliary field $D$, and a chiral scalar field $\Phi$.
The lowest component of $\Phi$ is a complex scalar
$\phi=(\Sigma+iA_5)/\sqrt{2}$, where $A_5$ is the fifth component of the
vector field.
The five-dimensional Lagrangian density is given by
\begin{equation}
\left[\frac{1}{g^2}\left(\Phi^{\dagger}\Phi
-\sqrt{2}(\Phi^{\dagger}+\Phi)\partial_y V -V\partial_y^2 V
\right)\right]_{\theta^4}
+\left[W_\alpha W^\alpha \right]_{\theta^2}+h.c.
\end{equation}
We assume a Fayet-Iliopoulos term on a brane at $y=0$, which looks like
\begin{equation}
\left[2\xi V \delta(y)\right]_{\theta^4},
\end{equation}
and matter fields $\phi_X^{\pm}$ with charges of $\pm q_X$ localized  on a 
brane at $y=L/2$. 
The D-flat condision is
\begin{equation}
-D=\left[2 \xi \delta(y) +\frac{g q_X}{2} (|\phi_X^{+}|^2 -
|\phi_X^{-}|^2)\delta(y-L/2) + \partial_y \Sigma \right]=0
\end{equation}
which is satisfied although the fields  $\phi_X^{\pm}$ are not 
located at the brane where the the Fayet-Iliopoulos term is localized.
The explicit form of the solution is 
$|\phi_X^{+}|^2 -|\phi_X^{-}|^2=-4\xi/g q_X$, $\Sigma=\xi
\epsilon(y)$.
This simple example shows that there is a possibility that supersymmetry 
can be restored by the derivative terms even if the 
Fayet-Iliopoulos term and the charged matter field are separated.
Such a configuration is possible for five-dimensional models with
orbifolded boundary conditions, at least when the Fayet-Iliopoulos term
and the charged matter are located exactly at the different fixed points.
On the other hand, if either or both of the branes
were displaced from the fixed points at the beginning of inflation,
there is no solution that satisfies both the D-flat condition and
the orbifold boundary condition for $\Sigma$.
Moreover, for the models with more than two extra dimensions,
it seems rather robust to expect that such non-trivial configurations 
always exist to compensate the supersymmetry breaking.

Thus our conclusion is the following.
For the simplest case with the restricted initial conditions, 
and with the help of the derivative terms,
one can find non-trivial configuration that
satisfies the D-flat condition, even if the Fayet-Iliopoulos terms and
the charged matter fields are located at a distance.
However, such an example seems rather peculiar for the models that we
have considered in this paper.
As a result, in our models for inflation, it is natural to expect that 
supersymmetry is
broken at the beginning of inflation, except for the specific models
with restricted initial conditions, where derivative terms are effective
to recover supersymmetry. 

The most peculiar point in our model is the suppression of the
interaction between inflaton and the source of the supersymmetry
breaking during inflation. 
In our model, the tree-level interaction between inflaton (moduli for
the distance between branes) and the fields on the branes appears only
through the exponential factor. 
One should recall that in the conventional models for hybrid inflaton
the trigger field must have ``large'' coupling to the inflaton field
in order to stabilize the trigger field on top of the potential.

\section{Cosmological constraints}
\hspace*{\parindent}
In this section we examine the cosmological constraints for the above
models.
When one considers inflation, one of the most obvious expectations will
be that it explains the origin of the cosmic microwave background (CMB)
anisotropy of the present Universe.
On the other hand, the requirement from the COBE data sometimes
imposes fatal constraint on the models for inflation.
Although the constraint may disappear if alternative mechanisms, such as
cosmic strings\cite{vilenkin-book} or curvaton
hypothesis\cite{curvaton} works well to meet all the observational
bounds, it is still very important to ask whether one can find a model
for inflation that produces the required density fluctuation.
In this section we examine whether the above model for inflation 
can produce the required anisotropy during inflation without
fine-tunings.

We think it is important to show why the conventional models for hybrid
inflation needed peculiar fine-tunings.
The requirement from the
COBE measurement puts severe bounds on their scales and couplings,
because of the large loop correction.
For example, we consider the original model for hybrid
inflation\cite{original} with the potential  
\begin{eqnarray}
V(\phi, \sigma)&=& V_0 + \frac{1}{2}m_{\sigma}^2 \sigma^2 + 
\frac{1}{2} g \phi^2 \sigma^2 \nonumber\\
&& \frac{1}{4}\lambda \phi^4 - \frac{1}{2} m_{\phi}^2 \phi^2.
\end{eqnarray}
The ``large'' loop correction comes from the $\phi$ field.
If supersymmetry remains, the result is simplified because only the
logarithmic part is relevant.
The form of the loop correction is
\begin{equation}
\Delta V_{1-loop}(\sigma)= \frac{1}{64 \pi^2}
\left( m^4(\sigma) ln \frac{m^2(\sigma)}{\Lambda^2} \right)
\end{equation}
where 
\begin{equation}
m^2(\sigma) = (g^2 \sigma^2 -m_{\phi}^2)
\end{equation}
and $\Lambda$ is the renormalization scale.
The flatness conditions require\cite{lyth}
\begin{equation}
g \ll \frac{<\sigma>}{M_p},
\end{equation}
which means that the non-renormalizable terms cannot be ignored 
if $g$ is not fine-tuned.
Moreover, the COBE normalization requirement gives an additional
constraint 
\begin{equation}
<\phi>^4 \sigma_{COBE} \, \ge\,  (10^9 GeV)^5
\frac{V_0^{\frac{1}{2}}}{(1 MeV)^2} 
\end{equation}
where $\sigma_{COBE}$ denotes the expectation value of $\sigma$ when the
COBE scales leave the horizon.
These conditions are crucial for the models with large extra
dimensions\cite{lyth}.

In our model, however, the situation is quite different.
The ``trigger'' mechanism is not due to the renormalizable couplings
between large inflaton field and the trigger field, but is induced by
the brane separation.
Thus there is no need for the large ``direct'' coupling between inflaton and
the trigger field.
The relevant couplings are suppressed by the
exponential factor when the brane distance is larger than the width of
the wavefunctions.\footnote{See eq.(\ref{1+2}). Of course there are
higher dimensional terms that is not effective in our model.}
Thus it is quite easy to find that 
the conventional loop corrections are tiny and irrelevant in our model.
Although the serious constraint from the loop correction does not appear
in our model, another problem still remains because of the limit
for $\sigma_{COBE}$.
Here we consider $\sigma \equiv M_*^2 r_1$, which is the inflaton that
parametrizes the brane distance. 
Assuming that the inflaton fluctuation is the origin of the structure of
the Universe, one will find the constraint
\begin{equation}
M_p^{-3} \frac{V_0^{3/2}}{V'} = 5.3 \times 10^{-4}.
\end{equation}
This implies that 
\begin{eqnarray}
\label{sigma_cobe}
\sigma_{COBE} &\sim& M_p^{-3}V_0 ^{3/2} (5.3 \times 10^{-4})^{-1}
m_{\sigma}^{-2}\nonumber\\
&\sim& 10^{-4} GeV \left(\frac{V_0}{(10^{5}GeV)^4}\right)^{3/2}
\left(\frac{V_0/M_p^2}{m_\sigma^2}\right)
\end{eqnarray}
where $\sigma_{COBE}$ is the expectation value of the inflaton 
when scales explored by COBE leave the horizon.\footnote{We have assumed
the simplest form of the potential, $V\simeq V_0 + m_{\sigma}^2 \sigma^2$.}
Of course, $\sigma_{COBE}$ must not be smaller than $M_*$, where brane 
inflation ends.
Obviously, the bare mass for the $\sigma$ field is required to be
smaller than $\sqrt{V_0/M_p^2}$.
If supersymmetry is broken on the brane and the transition to the bulk
fields occurs at the tree level, one can estimate an {\bf upper} limit
for the soft mass by dimensional analysis\cite{Extra_1},
\begin{equation}
m^2_{modulus}\sim G_{4+n_{E}} \frac{|F_{brane}|^2}{R_E^{n_{E}}},
\end{equation}
where $G_{4+n_{E}}$ is the gravitational constant in the $4+n_{E}$
dimensions and $F_{brane}$ denotes the supersymmetry breaking on the
brane.
Without additional symmetries or mechanisms, the soft masses for the
modulus can be expected to be a few orders smaller than the above upper limit.
The lower limit for $F_{brane}$ must be given by the requirement from
the conventional 
soft supersymmetry breaking terms in the supersymmetric extension of the 
standard model, which cannot be much smaller than the TeV scale. 
Thus the most optimistic requirement is $F_{brane} \geq$O(1)TeV. 
In our model for D-term inflation, we are considering inflation
where supersymmetry breaking is dominated by the D-term.
We may safely assume that $|F_{brane}|^2 << V_0$ during
inflation. 
From the above arguments, we can see that eq.(\ref{sigma_cobe})
puts a serious constraint on F-term inflation, while it does not 
exclude D-term inflation with $M_* \simeq 10^{5-6}$GeV.
Unlike the conventional models for D-term inflation, 
no ``large'' interaction is required between inflaton and the trigger
field, which avoids the most serious constraint from the loop
corrections.

\underline{Forces between branes}

To calculate the cosmological parameters, we must first 
determine the form of the potential for the inflaton field.
This matter is already discussed by many authers\cite{angle_inflation}.
Here we stress that in our model we do {\it not} always assume that the 
trigger field (the source of the vacuum energy during inflation)
dominates the corrections that lift the flat inflaton potential.
When we consider D-term inflation, one may expect that the
phenomenological source of the supersymmetry breaking dominates the
force between branes. 
Of course, one may expect that many types of corrections are present 
at the same time, which take  
different forms.
For example,
\begin{itemize}
\item The simple $m_{3/2}^2$ correction from supergravity.
\item Since the cancellation between the graviton-dilaton attraction 
and the RR repulsion fails when supersymmetry is broken, the potential 
of the form\cite{thermal_brane}
\begin{equation}
\label{analogy}
V(\sigma)\simeq M^4\left(1-\frac{m^k}{\sigma^k}\right),
\end{equation}
must appear. Here $m$ and $k$ are not fixed in our setups for D-term
      inflation.
\item The loop corrections to the K\"ahler metric that comes from 
the particles of the mass $\phi=M r^2$\cite{thermal_brane}.
\end{itemize}
The simplest example with $m_{\sigma}^2 \simeq m_{3/2}^{2}$ is
already discussed.
In models where the Vam der Waals forces between branes dominates 
the potential, the flatness conditions on the potential is satisfied when
\begin{equation}
\epsilon = \frac{M_p^2}{2}\left(\frac{V'}{V}\right)^2
\simeq \frac{M_p^2}{2}\left(\frac{m^k}{\sigma^{k+1}}\right)^2 << 1
\end{equation}
and
\begin{equation}
|\eta|=\left| M_p^2 \frac{V''}{V}\right|
\simeq 2M_p^2 \frac{m^k}{\sigma^{2+k}} << 1.
\end{equation}
One may expect that the one-loop corrections to the K\"ahler
potential lift the inter-brane potential\cite{thermal_brane}
 when branes are well separated,
\begin{equation}
\label{loopf}
V(r)\sim M^4 (1+c)ln(M_* r).
\end{equation}
In the case when eq.(\ref{loopf}) dominates the effective potential,
the situation fits to the original idea of D-term
inflation\cite{D-infla, D-matsuda}. 
The significant difference is that the serious constraint from the loop
corrections, which was suggested 
in ref.\cite{lyth}, does not exist in our model.
During inflation, the interaction between the trigger field and the
inflaton is not required to be large in our model. 

Our conclusion in this section is the followings.
For the F-term model, it is still difficult to make successful
inflation. 
On the other hand, for the D-term model, the unique problem of
the original hybrid inflation is safely removed. 
As we have discussed in the above,
the precise form of the inflaton potential
is not determined solely by the mechanism of non-tachyonic inflation, but
will rather be determined by the phenomenological model.

\section{Conclusions and Discussions}
\hspace*{\parindent}
We have studied inflation in models for the brane universe,
considering hybrid brane inflation without tachyon condensation.
In our model, when branes are far apart, supersymmetry is spontaneously 
broken by the
fields on a brane, which induces inflation.
The inflaton is the moduli for the brane distance.
At the end of inflation, when branes come close, supersymmetry
is restored by the interaction.
Then the field on the brane starts oscillation to reheat the Universe.
In this paper we have constructed explicit models for F-term and 
D-term
 inflation.
Although F-term inflation is not suitable for the first inflation,
D-term inflation works without any fine-tunings.
There are at least two major advantages in our model.
One is that the reheating is natural in our model,
because the trigger field is not the tachyon but a conventional field on
the brane. 
The serious constraint from the loop correction, which always appears 
when one considers conventional models for hybrid inflation
even if it is induced by the D-term, is removed
because the ``large'' coupling between inflaton and the trigger field is
not needed in our model.

\section{Acknowledgment}
We wish to thank K.Shima for encouragement, and our colleagues in
Tokyo University for their kind hospitality.

\end{document}